# Laboratories of Oligarchy?
# How the Iron law Extends to Peer Production

Aaron Shaw, Northwestern University
Benjamin Mako Hill, University of Washington

## 1. INTRODUCTION

Commons-based peer production – the distributed creation of freely accessible information resources through the mass aggregation of many contributions – represents a modality of collective intelligence that integrates the use of digital communication networks and information technologies [Benkler 2006; Benkler et al. 2013]. Peer production projects like Wikipedia and Linux have inspired numerous organizations, social movements, and scholars to embrace open online collaboration as a model of democratic organization with broad democratizing potential [Bennett and Segerberg 2013; Castells 2012; Shirky 2008]. However, many peer production projects exhibit entrenched leadership and deep inequalities [Healy and Schussman 2003; Ortega 2009; Viégas et al. 2007], and often explicitly embrace undemocratic organizational forms [Hill et al. 2008]. Instead of functioning as digital "laboratories of democracy," peer production projects may conform to Robert Michels' "iron law of oligarchy" ([1915]), an influential sociological theory which proposes that as membership organizations become large and complex, a small group of early members consolidate and exercise a monopoly of power within the organization as their interests diverge from those of the collective.

In a working paper that is currently under review,[1] we offer an empirical test of whether or not large peer production projects resist increasing levels of organizational oligarchy. Using exhaustive longitudinal data of internal processes drawn from 683 wikis that have grown large and complex, we adapt Michels' iron law to the context of peer production communities and construct a series of hypothesis tests. In contrast to previous empirical studies of peer production [Konieczny 2009] and much theoretical work, we present quantitative evidence in support of Michels' iron law.

## 2. BACKGROUND

Little previous empirical research on peer production has formally tested the pervasive claims about projects' democratic organizational governance [Benkler et al. 2013; Crowston et al. 2010]. For our purposes, organizational democracy consists in the active participation of multiple constituencies in the negotiation and exercise of legitimate authority within an organization. This definition derives from previous research on voluntary, membership, and movement organizations engaged in collective action of various kinds [Andrews et al. 2010; Lipset et al. 1956; Ostrom 1990; Voss and Sherman 2000] and builds on foundational studies of public goods production and common pool resource management [Hardin 1968; Olson 1965; Ostrom 1990].

As with other translations of Michels' iron law to organizations beyond political parties, oligarchy and democracy have distinct meanings in the context of peer production communities and we adapt and operationalize the theory to suit our empirical setting. Michels proposes that two patterns drive increasing oligarchy within voluntary organizations as they grow: (1) structurally, the group becomes an increasingly complex organization with a small group of professional leaders who exercise a monopoly over the mechanisms of authority; and that (2) these leaders develop independent interests in the preservation of the organization itself, resulting in the transformation of the goals and activities of the organization in ways that diverge from the interests of members. We refer to these distinct dynamics as the "structural" and "goal transformation" components of the iron law [Jenkins 1977; Leach 2005; Voss and Sherman 2000].

---

[1]http://mako.cc/academic/shaw_hill-laboratories_of_oligarchy-DRAFT.pdf

Previous studies of movement and voluntary organizations guide our operationalization of Michels' two components of oligarchy in the context of peer production projects. Structurally, the governance and leadership of oligarchic organizations must reside in the hands of a stable, entrenched, minority that exercises dominant control over organizational resources and policy [Lipset et al. 1956]. Oligarchic goal transformation occurs if organizational leaders and elites develop interests that diverge from those of other members and impose their interests on the rest of the organization. If it applies, Michels' iron law requires that both patterns hold across peer production projects as they grow over time.

## 3. RESEARCH DESIGN

Our empirical setting is a large sample of peer production communities engaged in the collaborative creation of wikis hosted by the for-profit firm *Wikia*. The term "wiki" refers to software designed to facilitate the collaborative, asynchronous creation and distribution of textual content, as well as the communities that use wiki software, and the products they create [Leuf and Cunningham 2001]. Wikipedia is the most famous wiki, but there are hundreds of thousands of other wikis with different goals, topics, and scopes. Several aspects make Wikia an ideal setting in which to analyze organizational governance. Founded in 2004, Wikia sought to apply the Wikipedia model of peer production beyond the education-based scope of the Wikimedia Foundation. Although many firms host wikis (e.g., PBWiki, WikiSpaces, and SocialText), Wikia is unique in that it only hosts publicly accessible, volunteer-produced, peer production projects. Like Wikipedia, anybody can create an account on any Wikia wiki. These factors ensure that Wikia wikis remain open and accessible, allowing communities to adopt more or less participatory and democratic organizational behavior.

We analyze the full records of intra-organizational behavior within 683 of the largest Wikia wikis. Our dataset includes every revision, from both registered and non-registered users, from the time of Wikia's founding in 2004 until the point of data collection in April, 2010. We focus our analysis on the activity of wiki administrators – the individuals within the projects in formal positions of authority. Using hierarchical linear models with random intercepts as a multilevel model for change, we test three hypotheses that correspond to the "structural" and "goal transformation" components of the iron law: **H1:** The probability of adding new administrators declines as wikis grow (in terms of the number of contributors). **H2:** Controlling for total administrative activity, administrators will engage in more administrative activity as wikis grow. **H3:** Controlling for contributions by experienced contributors, the number of contributions from experienced contributors that are reverted (deleted) by administrators will increase as wikis grow.

## 4. RESULTS

The results of the three fitted models (as well as additional robustness checks) are available in the working paper, and we summarize them briefly here. With respect to H1, we find that, *ceteris paribus*, one log-unit increase in the number of registered contributors is associated with odds of adding a new administrator which are only 0.81 times as high. For H2, we estimate that a 1% change in the total number of accounts on a wiki, controlling for the total edits made to administrative spaces of the wikis, is associated with a 3% increase in the number of administrative contributions made by administrators. Our models also support H3, and at the margin, we estimate that a 1% increase in the total number of accounts on a wiki, controlling for the total number of edits by experienced users in the wiki that week, is associated with a 5% increase in the number of administrator-reverted contributions made by experienced editors.

Plots of predicted values for prototypical wikis for each of our three models are shown in Figure 1 along with 95% confidence intervals for the marginal effects. These model-derived plots emphasize what happens, on average, in our sample. The values along the $y$-axis correspond to estimated values of each of our dependent variables in our three models. Along the $x$-axis of each plot are a range of likely values of total registered accounts from 0 to 793 (the 95$^{th}$ percentile of observations in our dataset). These plots show the predicted values for wikis that differ in terms of the number of registered users with at least one edit but are identical in every other respect. We have held each of our control variables constant at the sample median. Each of these prototypical values should be interpreted in the context of a single week-long period.

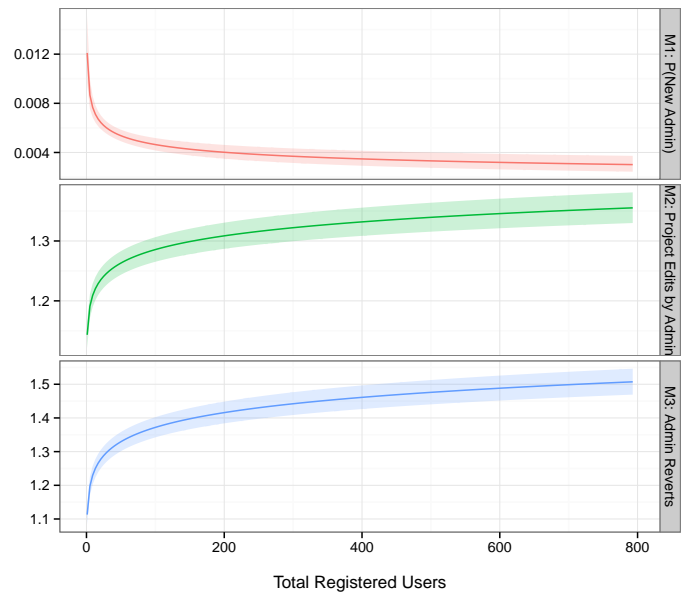

Fig. 1. Predicted values (including 95% confidence intervals) from our models for wikis with varying numbers of accounts holding all other variables at sample medians. All outcome variables measured in "per week" units.

## 5. CONCLUSION

We find strong evidence that, on average, as wikis become larger, a small group – present at the beginning – monopolizes positions of formal authority in the community and accounts for more administrative activity while also using their authority to restrict contributions from experienced community members. The wikis in our sample are not indicative of robustly democratic, participatory institutions. This is true despite the relative lack of formal bureaucratic structure or clearly-defined roles within many wikis. These results are consistent with Michels' iron law of oligarchy and contradict prevailing theoretical and empirical findings regarding organizational democracy in peer production. By conducting a comparative analysis across peer production communities, we also contribute a large-scale empirical analysis of the organization of collective intelligence systems.

We note that some wikis in our dataset appear more robustly democratic than others. Understanding why *some* peer production projects and collective intelligence systems create robust democratic organizations is a promising area for future research. Nevertheless, our results imply that widespread efforts to appropriate online organizational tactics from peer production may facilitate the creation of entrenched oligarchies in which the self-selecting and early-adopting few assert their authority to lead in the context of movements without clearly defined institutions or boundaries. The widely observed hierarchies and inequalities within peer production projects and collective intelligence systems also bring enhanced oligarchic organizational structures and behavior among project leaders.

## REFERENCES

Kenneth T. Andrews, Marshall Ganz, Matthew Baggetta, Hahrie Han, and Chaeyoon Lim. 2010. Leadership, Membership, and Voice: Civic Associations That Work. *Amer. J. Sociology* 115, 4 (2010), 1191–1242. DOI:http://dx.doi.org/10.1086/649060

Yochai Benkler. 2006. *The Wealth of Networks: How Social Production Transforms Markets and Freedom*. Yale University Press.

Yochai Benkler, Aaron Shaw, and Benjamin Mako Hill. 2013. *Peer production: A modality of Collective Intelligence*. Technical Report. Unpublished manuscript.

W. Lance Bennett and Alexandra Segerberg. 2013. *The Logic of Connective Action: Digital Media and the Personalization of Contentious Politics*. Cambridge University Press.

Manuel Castells. 2012. *Networks of Outrage and Hope: Social Movements in the Internet Age*. John Wiley & Sons.

Kevin Crowston, Kangning Wei, James Howison, and Andrea Wiggins. 2010. Free/libre open source software: What we know and what we do not know. *Comput. Surveys* 44, 2 (2010), 2012.

Garrett Hardin. 1968. The Tragedy of the Commons. *Science* 162, 3859 (1968), 1243–1248. DOI:http://dx.doi.org/10.1126/science.162.3859.1243 PMID: 17756331.

Kieran Healy and Alan Schussman. 2003. The ecology of open-source software development. (2003).

Benjamin Mako Hill, Corey Burger, Jonathan Jesse, and Jono Bacon. 2008. *Official Ubuntu Book* (3 ed.). Prentice Hall.

J. Craig Jenkins. 1977. Radical Transformation of Organizational Goals. *Administrative Science Quarterly* 22, 4 (1977), 568–586. DOI:http://dx.doi.org/10.2307/2392401

Piotr Konieczny. 2009. Governance, Organization, and Democracy on the Internet: The Iron Law and the Evolution of Wikipedia. *Sociological Forum* 24, 1 (2009), 162–192. DOI:http://dx.doi.org/10.1111/j.1573-7861.2008.01090.x

Darcy K Leach. 2005. The Iron Law of What Again? Conceptualizing Oligarchy Across Organizational Forms. *Sociological Theory* 23, 3 (2005), 312–337. DOI:http://dx.doi.org/10.1111/j.0735-2751.2005.00256.x

Bo Leuf and W. Cunningham. 2001. *The Wiki way: quick collaboration on the Web*. Addison-Wesley Longman Publishing Co., Inc. Boston, MA, USA.

Seymour Martin Lipset, Martin A. Trow, and James S. Coleman. 1956. *Union democracy: The Internal Politics of the International Typographical Union*. Free Press, Glencoe, Ill.

Robert Michels. 1915. *Political Parties: A Sociological Study of the Oligarchical Tendencies of Modern Democracy*. Hearst, New York, NY.

Mancur Olson. 1965. *The Logic of Collective Action: Public Goods and the Theory of Groups* (first edition ed.). Harvard University Press.

Felipe Ortega. 2009. *Wikipedia: A Quantitative Analysis*. PhD dissertation. Universidad Rey Juan Carlos.

Elinor Ostrom. 1990. *Governing the Commons: The Evolution of Institutions for Collective Action*. Cambridge University Press, New York, NY.

Clay Shirky. 2008. *Here Comes Everybody: The Power of Organizing Without Organizations*. Penguin Press HC, The.

Fernanda B. Viégas, Martin Wattenberg, Matthew M. McKeon, and D Schuler. 2007. The hidden order of Wikipedia. In *Online Communities and Social Computing*. Springer-Verlag, Berlin, Heidelberg, 445–454.

Kim Voss and Rachel Sherman. 2000. Breaking the Iron Law of Oligarchy: Union Revitalization in the American Labor Movement. *Amer. J. Sociology* 106, 2 (2000), 303–349. DOI:http://dx.doi.org/10.1086/ajs.2000.106.issue-2 ArticleType: research-article / Full publication date: September 2000 / Copyright © 2000 The University of Chicago Press.